\documentstyle[aps,pra,amsfonts]{revtex}
\tightenlines
\newcommand{\be}{\begin{equation}}
\newcommand{\ee}{\end{equation}}
\newcommand{\ben}{\begin{eqnarray}}
\newcommand{\een}{\end{eqnarray}}
\newcommand{\bF}{\begin{figure}}
\newcommand{\eF}{\end{figure}}
\title{Bohmian picture of Rydberg atoms}
\author{Partha Ghose and Manoj K. Samal}
\address{S. N. Bose National Centre for Basic Sciences,
Block JD, Sector III, Salt Lake, Kolkata 700 098}
\author{Animesh Datta}
\address{Department of Electrical Engineering, Indian Institute
of Technology, Kanpur, 208 016}
\begin{document}
\maketitle
\begin{abstract}
Unlike the previous theoretical results based on standard quantum
mechanics that has established the nearly elliptical shapes for the
centre-of-mass motion using numerical simulations, we show analytically
that the Bohmian trajectories in Rydberg atoms are nearly elliptical.
\end{abstract}

\section{Introduction}
Ever since the advent of quantum mechanics one of the fundamental
problems in physics has been to elucidate the transition from
quantum to classical mechanics. After Schr\"odinger introduced
coherent states in quantum mechanics \cite{schrod}, it became
clear that they are optimal quantum mechanical states to describe
the classical limit. Recently, there has been a renewed interest
in this problem due to the experimental study of Rydberg atoms in
external fields which provide a tool to study the
quantum-classical transition regime \cite{stebbings,alber}. These
experiments\cite{stroud1} have shown that high quantum number
states have the property of showing classical behaviour. Also,
higher $n$ values lead to an ever-shrinking energy gap $E_n =
-1/{2 n^2}$, and the energy spectrum approaches the continuum.
Moreover, it is evident \cite{nauenberg} that the highest angular
momentum state $l=n-1=l_n$ in an atom has the special property of
being a minimum uncertainty state.

In this paper, we study the problem from the point of view of the
de Broglie-Bohm theory \cite{bohm} to see the relationship between
Bohmian trajectories and Kepler orbits in Rydberg atoms.

\section{Standard quantum mechanical analysis}
It has been shown that wave-packet solutions of the Schr\"odinger
equation for the Coulomb problem travel along classical elliptic
orbits of fixed mean eccentricity and angular momentum
\cite{nauenberg}. These wave-packets are coherent states that have
minimal quantum fluctuations in the non-commuting components of
the Runge-Lenz vector in the plane of the orbit\cite{runge}. The
study of the Runge-Lenz vector in quantum mechanics was first
undertaken by Pauli\cite{pauli} who used it to solve the Hydrogen
atom problem.

In the asymptotic regime of large quantum numbers $n \approx l_0$,
the WKB approximation of the radial wave-function of the hydrogen
atom in $3$-dimensions is given by\cite{nauenberg}

\ben \psi^\delta(r, \theta, \phi, t) &\approx& \Big(\frac{2
\omega_0}{\pi p_0(r)}\Big)^{1/2} \exp{\Big[i S_0(r)\Big]}
\Big(\frac{\delta l_0}{\pi}\Big)^{1/4} \exp{\Big[-(\theta -
\pi/2)^2 \frac{\delta l_0}2{}\Big]} \nonumber\\& \times&
\sum_{\mu=- \infty}^{+ \infty} \exp{\Big[i \delta l_0 (\phi + 2
\pi \mu) - (\phi + 2 \pi \mu - \phi_0(r))^2 \frac{l_0}{2}
(1-\delta^2)\Big]} \nonumber\\& \times& \frac{1}{(2 \pi
\alpha_0(r,t))^{1/2}} \exp{\Big[\frac{- (\delta(\phi + 2 \pi \mu -
\phi_0(r))- \frac{t - t_0(r)}{l_0^3})^2}{2 \alpha_0(r,t)}\Big]}
\een where the classical action in radial coordinates is given by
\be
S_0(r)= \int^r p_0(r') d r' \ee with the mean radial momentum
\be
p_0(r) = \Big(2 E_{n_0} - \frac{l_0^2}{r^2} + \frac{2}{r}
\Big)^{1/2}. \ee In classical mechanics the quantities $\phi_0(r)$
and $t_0(r)$ are related to the action $S_0$ by
\be
\phi_0(r)= - \frac{\partial S_0}{\partial l_0} \:\: {\rm and} \:\:
t_0(r) = \frac{\partial S_0}{\partial E_{n_0}}. \ee The rate of
spreading of the wave-packet is given by the complex width
\be
\alpha_0(r, t, \delta) = \frac{1}{2 \sigma^2} - i \Big(\frac{3
t}{l_0^4} + 2 f_0(r) \Big)\ee where $f_0(r)={\partial^2 S
}/{\partial E_n^2}$. It follows from eqn. (1) that the quantum
mechanical phase is
\be
S= S_0 + \delta l_0 \phi + \frac{1}{2} \arctan (b/a) +
\frac{\lambda^2}{4 \gamma} - \frac{A^2 b}{2(a^2 +
b^2)}\label{action}\ee where \be A = \delta (\phi -\phi_0(r)) -
\frac{1}{l_0^3} (t - t_0(r)), \:\:\: \gamma = \frac{2 \pi^2 b
\delta^2}{(a^2 + b^2)}, \:\:\: \lambda = \frac{\pi \delta A}{(a^2
+ b^2)}\ \ee with
\be
a = \frac{1}{2 \sigma^2} \:\:\: {\rm and} \:\:\: b = \frac{3
t}{l_0^4} + 2 f_0(r). \ee

\section{Bohmian mechanical analysis}

Let us now examine the problem of coherent states from the
perspective of the de-Broglie-Bohm theory which is quite close to
the classical Hamilton-Jacobi theory. In this theory the quantum
mechanical action $S$ given by (\ref{action}) satisfies the
modified Hamilton-Jacobi equation
\be
\frac{\partial S}{\partial t} + \frac{(\nabla S)^2}{2 m} + Q + V =
0 \ee where
\be
Q = -\frac{\hbar^2}{2m} \frac{\nabla^2 R}{R}\ee and the
wavefunction, written in the polar form $\psi = R e^{i S/\hbar}$,
satisfies the Scr\"odinger equation
\begin{equation}
i\hbar \frac{\partial \psi}{\partial t}= [-\frac{\hbar^2}{2m}
\nabla^2 + V(x)] \psi
\end{equation}

In Bohmian mechanics one introduces the position as an additional
variable (the so-called ``hidden variable'') through the guidance
conditions\cite{holland}
\be
v_r = \frac{\partial S}{\partial r}, \:\: v_\theta = \frac{1}{r}
\frac{\partial S}{\partial \theta}, \:\: v_\phi = \frac{1}{r
\sin{\theta}} \frac{\partial S}{\partial \phi}\label{guid} \ee
Using these conditions, we can readily obtain the Bohmian
trajectories of electrons in a Rydberg atom with high $l$ wave
packets provided we also use the result that the radius of the
$n$th Bohr orbit in a hydrogen atom is given by $r_n \approx n^2$.
This is the expectation of the radius or its space averaged value.
The time averaged value is the same. Using the expression for $S$
given by (\ref{action}), we finally obtain

\ben \frac{\partial S}{\partial r}&=& p_0 + {\cal O}(l_0^{-7})+
{\cal O}(l_0^{-14})\label{s1} \\\nonumber \frac{\partial
S}{\partial \theta}&=& 0 \label{s2} \\\nonumber \frac{\partial
S}{\partial \phi}&=& \delta l_0 + {\cal O}(l_0^{-5})\label{s3}
\een The corresponding expressions in classical mechanics are \ben
\frac{\partial S}{\partial r}&=& p_0
\\\nonumber
\frac{\partial S}{\partial \phi}&=& \delta l_0 \\\nonumber \een
and these are known to lead to the equation of an
ellipse\cite{goldstein}. Eqns. (\ref{s1}) and (\ref{s3}) show that
the Bohmian trajectories of Rydberg atoms with $l_0 \approx
50-100$ are ellipses for all practical purposes.

\section{Concluding Remarks}

The advantage of the Bohmian analysis we have carried out is that
one can see clearly and analytically that the Bohmian trajectories
in Rydberg atoms are nearly elliptical. Previous theoretical
results using standard quantum mechanics have only been able to
establish the nearly elliptical shapes for the centre-of-mass
motion using numerical simulations \cite{stroud2}.

\section{Acknowledgements}

PG and MKS acknowledge financial support from the Department of
Science \& Technology, Govt. of India for this work. AD is
grateful to the S. N. Bose National Centre for Basic Sciences for
providing facilities for a summer project that enabled this work
to be undertaken.

\end{document}